\documentclass[12pt]{amsart}
\usepackage[english]{babel}

\usepackage{amsmath}
\usepackage{amssymb}
\usepackage{amsthm}
\usepackage{url}
\usepackage{graphicx}
\usepackage{algorithmicx,algpseudocode}
\usepackage[section]{algorithm}
\usepackage{caption}
\usepackage{wrapfig}
\usepackage{subcaption}
\newtheorem{theorem}{Theorem}[section]

\theoremstyle{remark}

\newtheorem{question}[theorem]{Question}
\usepackage{csquotes}

\DeclareMathSymbol{\minus} {\mathord}{operators}{"2D}

\newcommand\Var{\textrm{Var}}

\usepackage[colorlinks=true, allcolors=blue]{hyperref}

\title{What is the Sharpe Ratio, and how can everyone get it wrong?}
\author{Igor Rivin}
\address{Mathematics Department, Temple University and The Cryptos Fund}
\email{rivin@temple.edu}

\begin{document}
\begin{abstract}
The Sharpe ratio is the most widely used risk metric in the quantitative finance community - amazingly, essentially everyone gets it wrong. In this note, we will make a quixotic effort to rectify the situation.
\end{abstract}
\maketitle

\section{Introduction}

The Sharpe ratio was introduced over fifty years ago by William F.~Sharpe in \cite{sharpe1966mutual} (Sharpe revised the definition slightly almost thirty years later in \cite{sharpe1994sharpe}). The Sharpe ratio is a measure of risk-adjusted returns, and was initially intended to distinguish truly superior strategies from ones where the portfolio manager simply levered up a mediocre strategy. Such levering up would outperform the market in good times, and then crash in burn when things turned bad. The Sharpe ratio (in its most current incarnation) is defined as:

\begin{equation}
\label{sharpeq}
S_a = \frac{E(R_a - R_b)}{\sigma_a},
\end{equation}
where $R_a$ is the expected return of the asset, $R_b$ is the risk-free return rate, and $\sigma_a$ is the standard deviation of the asset returns. A statistically inclined reader will be sure to note the more-than-passing resemblance of the Sharpe ratio to the $t$-statistic, and so quantifies the evidence that the investment strategy is better than the proverbial monkeys throwing darts at the (by now virtual) stock table.

While the Sharpe ratio is widely used by investors (to decide which investment vehicle is preferable), it is also very significant as an internal measure in the quantitative finance community - a high Sharpe ratio indicates that it is extremely unlikely that the strategy will lose money in any given year, and so allows the portfolio manager to lever up the portfolio without too much risk of ruin.

\section{How is the Sharpe ratio computed?}
The quantities in formula \eqref{sharpeq} are \emph{annual} quantities, and so, in principle, to get reasonable estimates of all the variable, we should examine a portfolio manager's return over a few decades. This is obviously impractical - the portfolio manager may well be retired (or dead) by the time a reasonable value is obtain, and the value will be meaningless in any case, since the character of the market changes considerably over such lengthy time scales. Consequently, in practice, Sharpe ratio is computed using the \emph{daily} (sometimes monthly)  return stream. In order to make this computation feasible, a basic assumption is made:
\begin{quotation}
The daily returns are independent identically distributed (i.i.d) random variables.
\end{quotation}
While this assumption is clearly false on a number of grounds (there is seasonality in the markets, so the returns are not identically distributed, and there are momentum and reversal phenomena, which means that they are not independent), these assumptions are not too far from reality, and we will not quarrel with them here, since the real confusion is just beginning:

An assumption is made that (since the returns are small) the return over a number of days equals \emph{sum} of the returns on the individual days, or
\[
\prod_{i=1}^n (1+r_i) = 1+ \sum_{i=1}^n r_i.
\]
This is then used in the following way: the yearly return is the sum of 252 (the traditionally accepted number of trading days in a year) daily returns. Since the mean and variance are both expectations, both grow linearly with the size of sample. The standard deviation is the square root of variance, so, when the smoke clears, the formula used universally is:

\begin{equation}
\label{dailysharpe}
S_a = \sqrt{252} \dfrac{E(R_a(d) - R_b(d)}{\sigma_a(d)},
\end{equation}
where the $d$ now indicates \emph{daily} returns.

\section{The correct way}
Let us now see what the truth is.

As mentioned above, the actual annual return over $n$ periods is
\[
X_n = \prod_{i=1}^n(1+r_i).
\]
Since the daily returns are i.i.d, the expectation of the product is the product of the expectations, so 

\[
E(X_n) = (1+ \mu(d))^n.
\]

What about the variance?

\begin{gather*}
\Var(X_n) = E(X_n^2) - E(X_n)^2 = \\ E(\prod_i=1^n(1+r_i)^2] - (1+\mu(d))^{2n} = \\\prod_{i=1}^nE(1+r_i)^2) - (1+\mu(d))^{2n} = \prod_{i=1}^n ((1+\mu)^2) + \sigma^2) - \prod_{i=1}^n(1+\mu)^2 = \\
\sum_{i=1}^n \binom{n}{i}\sigma^{2i} (1+\mu)^{2n-2i}
\end{gather*}

So, finally, we have the following formula for the Sharpe ratio under the identical independent daily returns assumption:

\begin{center}
\boxed{
I_a = \dfrac{(1+\mu)^n-1}{\sqrt{\sum_{i=1}^n \binom{n}{i}\sigma^{2i} (1+\mu)^{2n-2i}}},}
\end{center}

Where $\mu, \sigma$ is the mean and standard deviation (respectively) of daily returns. The first question is:
\begin{question} is $I_a$ close to $S_a$ under the assumption of small returns?
\end{question}
The answer is: \textbf{NO.} Indeed, if the returns are quite small, then it is not unreasonable to approximate the numerator of $I_a$ by $n \mu.$ If the volatility is also quite low, it is quite reasonable to say that only the first term in the sum in the denominator contributes significantly. When the smoke clears, we get the following approximation:

\begin{quotation}
For $\mu, \sigma \ll 1,$ \[I_a \approx \sqrt{n}\dfrac{\mu}{(1+(n-1)\mu)\sigma}.\]
\end{quotation}
Notice that this differs from $S_a$ by a factor of $(1+(n-1)\mu).$  Since $(n-1)\mu$ is approximately the yearly return (under our hypothesis) the error is nontrivial even under the small returns and volatilities hypothesis.

The errors are much more egregious a little further away from the "heat death" limit. Consider, for example, the performance of crypto-currencies. Bitcoin has run up hugely over the last few years, and if the usual formula $S_a$ is used, the (one year lookback) Sharpe of Bitcoin (at the time of writing), is around $2.5$ (see, for example \url{https://www.sifrdata.com/cryptocurrency-sharpe-ratios/}. By contrast, the \emph{correct} Sharpe ratio (as posted on \url{http://cci30.com}) is $0.83,$ which is not so different from the S\&P 500 over the same period.

\section{Why is the standard calculation so wrong?}
As far as this author can tell, the reason is more sociological than mathematical. Remember that the primary audience for Sharpe consisted of investment managers - people who were probably good salesmen with good connections, but no understanding of mathematics. In particular, no understanding of logarithms. So, for them, life became much simpler if $\log(1+x) = x,$ as a direct consequence, the product of returns is roughly the sum of returns. Before we heap derision on these people, we should note that from the viewpoint of Kelly betting (see the canonical reference \cite{maclean2011kelly}) we \emph{want} to deal with the logs, and so a very reasonable quantity to use is Sharpe ratio in log space, computed as:
\[
L_a = \sqrt{n} \dfrac{E(\log R_a(d) - \log(R_b(d)))}{\sigma(\log R_a(d)}.
\]
Now, $L_a$ is much closer to $S_a$ than to $I_a,$ and arguably it is a more reasonable risk measure: Suppose you have an investment that (with equal probability) multiplies by 8 or halves by 2. The mean return is $3.25,$ the variance of returns is $38.6875.$ The "correct Sharpe ratio" as computed by $I_a$ is $1.348 10^{-72},$ so it judges this a rather poor investment.By contrast, the log-sharpe $L_a$ ratio is $5.29,$ indicating that this is a very good investment. The regular sharpe ratio $S_a$ is $8.29.$ It is clear that the last two numbers are more indicative than the first one. Why is this happening? Because the the fluctuations in the tail end of the year (by which time the account holder almost certainly owns this, and all other, universes) dwarf the returns for \emph{most} of the year.

\section{Conclusion} It is this author's strongly held opinion that the Log-Sharpe ratio $L_a$ is the right metric to use. However, if you do want to compute the actual Sharpe ratio (and many portfolio managers are required to do so by their investors), then use $I_a.$
\bibliographystyle{alpha}
\bibliography{sample}

\end{document}